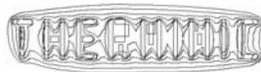

# High-Performance Thermal Interface Technology Overview

R. Linderman, T. Brunschwiler, B. Smith, and B. Michel
IBM Research GmbH, Zurich Research Laboratory
Säumerstrasse 4
8803 Rüschlikon, Switzerland

*Abstract*- **An overview on recent developments in thermal interfaces is given with a focus on a novel thermal interface technology that allows the formation of 2-3 times thinner bondlines with strongly improved thermal properties at lower assembly pressures. This is achieved using nested hierarchical surface channels to control the particle stacking with highly particle-filled materials. Reliability testing with thermal cycling has also demonstrated a decrease in thermal resistance after extended times with longer overall lifetime compared to a flat interface.**

I. INTRODUCTION

With the growing demand for integration density and faster electronics, the power density of chips continues to increase. Similar trends are also observed in processors, graphic chips, power electronic switches, light-emitting diodes, and microwave amplifiers. The power density problem thus affects a wide range of industries, from IC, power electronics, automotive, aerospace, defense, to medical. Thus there is an urgent need for more effective cooling to ensure performance gains and long- term reliability. Because thermal interfaces can consume up to 50% of the total thermal budget, a reduction of their resistance extends the lifespan of cooling solutions and helps to reduce junction temperatures. Packages with advanced cooling designs often use metal interfaces [1, 2], phase-change materials [3], or thermal interface materials made from highly particle-filled materials (greases and adhesives). Advantages of particle-filled materials are their lower cost, ease of assembly, and reworkability; disadvantages are a larger thermal resistance and the susceptibility to voiding and drying out for adhesives. Adhesives sacrifice re-workability [4] and high shear strength, but can be tailored to balance stresses between chip and substrate and can withstand relatively high cyclic strains.

Increasing the particle volume fraction exponentially increases the bulk thermal conductivity of a particle-filled polymeric materials because of percolation paths between particles in close contact [5]. As a drawback, the viscosity exponentially increases, preventing the formation of thin bond lines [6]. Newtonian and power-law fluid models can well describe purely viscous fluid materials. Particle-filled interface materials have been described by Bingham and Herschel–Bulkley fluid models, but measured bondlines are thicker than the models predict. Attempts to find the optimal size and volume fraction of filler particles to provide the lowest resistance use either viscosity [7] or mechanical viewpoints [8]. The latter assumes that thermal interfaces are pushed to particle contact and that packing determines the bondline thickness. One potential cause for voiding is repeated pressure and thermal cycles with air penetrating linked Hele-Shaw-type patterns [9]. While the physics of these systems is difficult to model, it is clear that pressure gradients, viscosity, and surface tension play a major role and voiding may be reduced by minimizing pressure gradients in thermal-grease bond lines during thermal cycles.

Solid particles in bi- or tri-modal size distributions are mixed into a viscous matrix at high volumetric loading levels—reaching 78% by volume to create thermal greases and adhesives [4]. Above the critical percolation threshold (~40 v% loading), the effective conductivity increases significantly owing to random networks of particles in contact with each other [10]. Commercial greases or pastes have effective bulk thermal conductivities ranging from 2–5 W/mK, with the thermal performance influenced mainly by particle loading and size distribution. Highly filled materials have nonlinear visco-elastic properties that can require squeeze loads high enough to damage flip-chip solder ball arrays and crack chips and substrates for larger chip sizes. In addition, an increase in bondline thermal resistance is observed because of the difficulty in achieving thin gaps close to the maximum particle sizes. Therefore, a trade-off is made between thin gaps with low-conductivity materials and thick gaps with high-conductivity materials [11]. Another reason why highly filled materials cannot be squeezed to thinner gaps is the stacking of particles during the squeezing process, which creates a rigid mechanical stop and a strong non-uniformity that can increase the likelihood of failure.

Many models and optimization techniques for TIM formulations have been proposed [4, 10-15]—all indicating improved bulk conductivity at higher particle loadings, yet without any indication of how to produce thin bondlines with more-highly-filled materials. The benefits of controlled particle orientations within the bondline have also been explored theoretically [15], but so far no technique to control particle orientation or position has been demonstrated. Prior efforts were motivated in part by the need to better understand the statistical non-uniformities associated with micro particle size and shape distributions and the complex rheology of highly filled materials. The hierarchical nested channel interface technology (HNC) represents a breakthrough in how to achieve thin bondlines with highly-particle-filled materials as it allows a reduction in assembly pressure, prevents particle stacking, and preserves material uniformity during bondline formation. As a result, an optimum TIM performance (see Fig. 1) can be achieved and micro- and nanoscale-engineered bondlines become feasible.

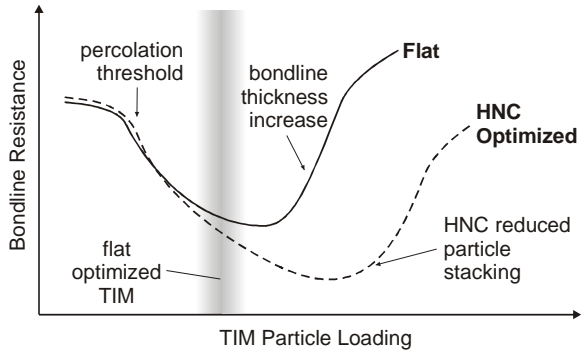

Fig. 1. Comparison of the bondline thermal resistance of conventional flat and HNC optimized bondlines.

## II. EXPERIMENTAL SETUP

The design of a test stand to evaluate thermal interfaces typically involves a trade-off between a system that closely simulates a microelectronic application and a system that allows quick and accurate evaluation of the interface. If the goal is to measure the effective TIM1 interface resistance in a planned product, modules with product chips are integrated with a lid and thermocouple between lid and cooler to estimate the junction-to-lid thermal resistance, where the interface resistance accounts for the largest part of the total. Interpretation of the results is complicated by the multi-dimensional heat flow due to spreading in the lid and often requires a correction factor determined by FEM simulation. However, testing large numbers of independent parameters can be impractical if multiple test vehicles must be built, and tested. These limitations lead to the development of research test stands allowing a more detailed measurement of properties in open and reusable systems integrating additional temperature sensors and the capability for in-situ bondline thickness measurement.

Figure 2a shows a test stand utilizing a liquid cooler and heater/sensor test chip similar to a microprocessor. To increase sensitivity to the thermal interface, a cold plate with minimum thermal resistance is utilized with a reference cold water supply temperature. Thermal interfaces typically represent over 60% of the total resistance for such a test stand for typical microprocessor package geometries and constraints. The main drawback of such a system is the uncertainty in the thermal resistance of the liquid cooler that is dependent on the liquid flow rate and temperature gradients in the direction of liquid flow. In addition, any localized hot spots in the heater can decrease the accuracy in the interface characterization. Typically, thermal resistance measurements are made at several thicknesses to extrapolate the effective conductivity of the interface material and to estimate the parasitic resistance of the cooler and chip. As the interface resistance is reduced below the uncertainty in temperature or parasitic resistance measurements, this style of test stand becomes impractical for further interface research. This typically occurs when the bondline thickness decreases below 5 µm or the interface has a resistance below 2 $Kmm^2/W$. To study higher conductivity materials or thinner gap interfaces, a new type of test system is required.

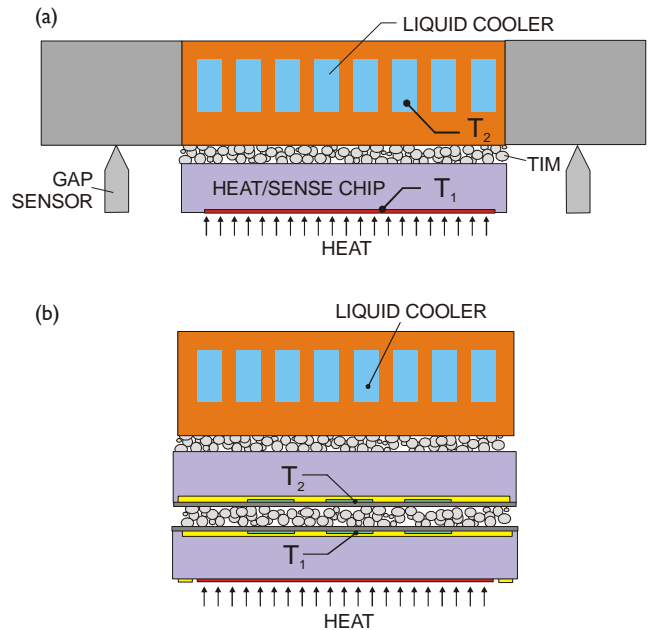

Fig. 2. (a) Conventional TIM test setup with heater chip and liquid cooler. The sensitivity of this technique depends on the TIM thickness and on the fact that the thermal resistance is a large portion of the total resistance and much larger than the measurement uncertainty. (b) Test stand integrating temperature sensing directly at the TIM layer.

While most techniques infer TIM properties by extensions of 1-D heat diffusion theory, transient methods can isolate the influence of the TIM from the other components in the thermal path [16, 17]. This reduces the inaccuracy in TIM properties due to uncertainty of the cooler properties, for example. The transient techniques introduce new uncertainties related to short-time data acquisition, power switching, and numerical model-fitting [18] and may not be well-suited for product-oriented test system.

A more radical departure from a product-oriented system that allows increased sensitivity to local interface properties is shown in Fig. 2b. The integration of temperature sensors directly on either side of the interface enables a direct measurement of the temperature drop across the interface under investigation as well as measurement of localized non-uniformities related to particle stacking or non-planar surfaces. Non-uniformities, however, can also complicate analysis of the measured data as they introduce ambiguity in the local heat flux. The test stand shown in Fig. 2b can also be used to study transient effects because one side of the interface can be driven with a power step or high-frequency signal. Sensitivity is limited only by the accuracy of the temperature measurement and the control of thin isolation layers that protect the sensor and wiring layer. A major challenge for ultrathin-bondline testing is the uncertainty in thickness created by non-planar chips or warpage from non-uniform support conditions.

### III. RESULTS

When a liquid is being squeezed between two parallel plates, a pressure-driven flow is generated. As with flow in a capillary, the velocity profile for a Newtonian fluid is parabolic across the channel. An approach to avoid thicker gaps is to sub-divide the chip or cooler surfaces into an array of protruding posts by means of a channel array with a depth $d$ that provides preferential flow paths for the flowing material (Fig. 3). Stefan's squeeze-flow theory predicts a narrower gap $H$ for squeezing viscous material between smaller protrusions. From the Hagen–Poiseuille law, we see that small and long capillaries offer a huge resistance to the fluid flow. A combination of the two theories provides an optimum post and channel size and an optimum number of subdivisions of the initial area to yield the smallest fluid resistance. A high thermal conductivity requires, in addition, a large area fill factor of the more conductive material and small channels between the protrusions. Because the flow resistance of fluidic networks scales with the inverse fourth power of the effective radius, channels have to be either short or wide and it is not possible to unify post size, channel width, and channel length in one set of channels. Instead, nested channels with increasing length and width must be created. These requirements are met when channel sizes are switched after a number of repeats in a self-similar manner. Such hierarchically nested channels (HNC) fulfill the conflicting requirements of all three parameters [19].

The first effect observed by increasing particle loading is the percolation threshold, at which the effective conductivity of the mixture rapidly increases above the 0.1–0.4 W/mK of the polymer matrix. At loadings near the stacking threshold, a large jump in bondline thickness occurs, which limits the usefulness of any gains made in effective bulk conductivity. Material segregation or filtering accompanied by increases in the required squeeze load have also been observed in prior studies [14, 20] when the squeeze velocity was below a certain critical velocity that depended inversely on the matrix viscosity. Linderman *et al.* [21] showed that the stacking threshold can be delayed by increasing the matrix viscosity. The delay in stacking with increased matrix viscosity is the result of stronger drag forces on particles and increased particle-settling times. Highly conductive materials made from a high-viscosity matrix and a high particle fill have an effective viscosity and yield strength that force nested channel surfaces to achieve thin bondlines with a reasonable assembly pressure. Particle-filled material with a higher matrix viscosity can be beneficial to extend the lifetime of thermal greases because of the slower drying, de-mixing and material pump-out associated with many thermal cycles.

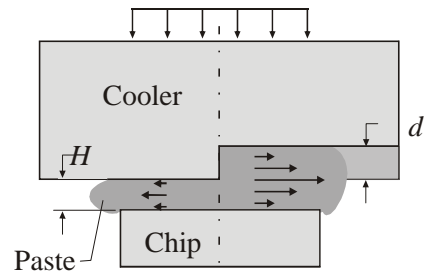

Fig. 3. Squeeze flow between two parallel plates (left) and with an additional channel (right). The arrows show the large flow difference due to the larger dimension of the channel plus the squeeze gap.

Visual inspection of the bondline through a glass chip often reveals two high-contrast lines between opposite chip corners, referred to here as the paste-cross (see Fig. 4, top). Despite having appeared in bondline inspections for many years [22], the origin and effects of the paste-cross were unclear. For some materials, the paste-cross is clearly visible to the eye, whereas other interface materials require scanning acoustic microscopy to identify the stacking region. The paste-cross forms because of the non-uniform pressure drop in the radial directions from the chip center created by the differences in flow lengths to the chip edge. As the interface material fills the entire chip area, a bifurcation in the flow develops along the paste-cross lines owing to the material flowing in a path of least resistance (arrows) towards the nearest chip edge rather than in a purely radial flow. Particles that lie on the bifurcation line during the squeeze flow are pulled in opposite directions by the viscous matrix. As a result they cannot move with the fluid and pile up, creating a mechanical obstruction of densely packed particles (see Fig. 4, right panel). A description of the velocity field present in the squeeze flow between square plates was given by Davidson *et al.* [14]. The stacking line initiates chip cracking and voiding in thermal greases under cyclic stress as well as mechanical failure in adhesive bonds (the stacking line lies along the region of highest shear stress in adhesive bonds having a thermal expansion mismatch). Higher particle concentrations in the paste-cross region can also exist in bondlines made with materials mixed below the stacking threshold when squeezed to thin gaps.

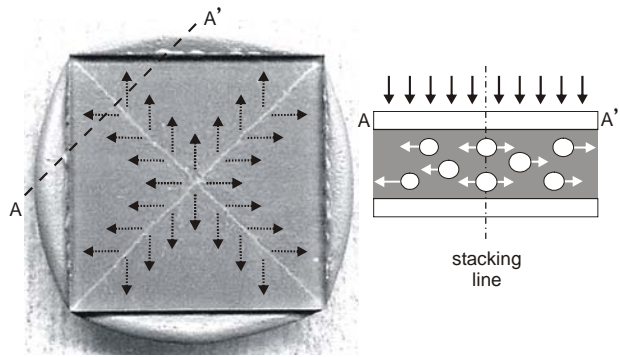

Fig. 4. "Paste-cross" as seen through a 14-mm glass chip with bifurcation in the flow pattern as interface material flows along the path of lowest pressure drop (left). Cross section along line A-A (right).

To redistribute the flow and prevent particle stacking, a nested surface channel array must be designed. Effective channel sizes useful for redistributing the flow can be estimated based on the TIM particle-size distribution and tests using a flat chip to determine at which bondline thickness particle stacking begins to limit further squeezing. Surface channels do not have a large impact on the flow until the bondline is thinner than a third of the channel depth—otherwise a much lower pressure drop exists through the remaining bondline thickness. The second important criterion is the largest particle size in the mixture, which defines the minimum possible bondline and thus the lower limit for channel sizes. Because the channels remove valuable surface area for heat conduction, they have to be as small as possible, dictating a hierarchical approach to channel design to optimize surface area and particle stacking. With an excessive number of HNC channels, the bondline thickness will continue to shrink, but the effective thermal resistance increases owing to the loss of surface area. One approach to hierarchical nesting is to define larger corner-corner channels to prevent non-uniformities from developing early in the bondline assembly and smaller-sized channel cells to manage stacking towards the end of bondline formation.

Figure 5 shows a particular HNC chip design with the resulting stacking pattern for a silver particle in PDMS test material. Note that the stacking lines bisect the interior angle of the channels of the HNC cells and intersect very close to the center of the triangular posts, indicating a proper channel design to control stacking. At interior nodes, the pressure drop to the chip edge is matched for each of the three intersecting channels and within the HNC cell.

The first hierarchy level is the larger corner-corner channels (black), the second is formed by the grey lines with the required hydraulic diameter enlargement for the diagonal channels (darker grey) as described by Linderman et al. [21]. A third level nesting can be made within the cells of the second level also using the previously defined pressure-drop-balancing methodology (not used in the design shown in Fig. 5). If additional hierarchy levels are patterned and the bondline cannot be squeezed to dimension where the finest channels redistribute the flow, then a particle stack line is directly inline with a channel, producing a poor thermal result because the channel will block the heat flow through the stack. The optimum cell size is determined by a series of tests within a range of sizes. One could make a trade-off between cell size and hierarchy level to further optimize the properties. In the examples presented here, a two-level hierarchy was used. Results for three-level hierarchies are not shown, as low-cost fabrication of caps with such structures was more difficult at the time of testing.

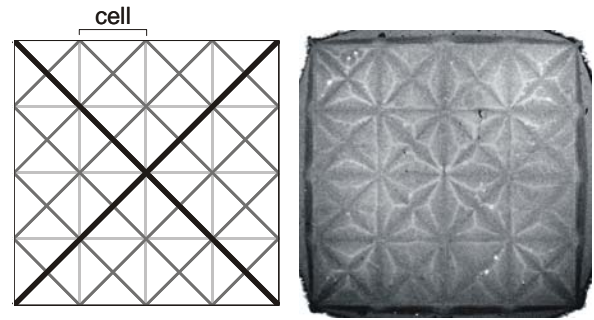

Fig. 5. HNC design (left) for a 14-mm chip with 3.5-mm-wide second-level HNC cells. Thick black lines represent the first-level hierarchy with 220 μm depth and width, grey lines the second level with 150 μm width and 180 μm depth, and dark grey lines its diagonal channels with 150 μm width and depth. The resulting particle-stacking pattern is shown on the right.

Figure 6 presents results for the IBM ATC thermal grease and for a Momentive Performance Material adhesive tested before curing. The thermal resistance of the HNC interfaces is lower because the surface channels enable thinner bondlines than a flat surface. Alternatively, the particle loading of the TIM could be increased to yield an improved conductivity at the same bondline thickness as can be achieved from flat surface with a lower-filled material. However, the TIM materials tested here were not altered from their original mixtures intended for flat surfaces and were used as supplied with the HNC surfaces.

Figure 6(a) shows that the resistance can increase quickly as the cell size becomes too small despite a thin bondline because of the loss in surface area when the cell pitch is too small. The added resistance of the HNC can be estimated by computing the reduction in cross-sectional area for solid heat conduction for a given channel design, with higher-conductivity materials, such as copper, supporting smaller cell pitches before the HNC resistance increases.

The HNC substrate reduces TIM thermal resistance by approximately a factor of two compared with that of a flat surface with the same assembly load due mostly to the corresponding reduction in bondline thickness by a similar amount. In product applications, the HNC will also be patterned on the copper chip cap, which for the H2-150-220 design will lead to an additional thermal benefit of 0.3 $Kmm^2/W$ compared with silicon values measured here.

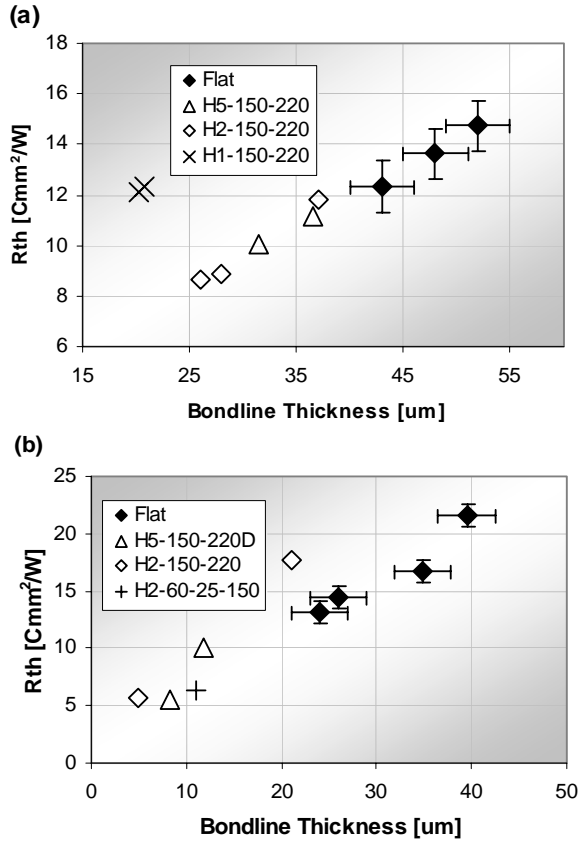

Fig. 6. Thermal resistance benefit from HNC interfaces compared with a flat surface. (a) IBM ATC D99 = 11 μm, and (b) Momentive Performance Material adhesive-2. HNCs with third-level 25-μm-deep channels (variable width) are marked by X. Bondline conditions: 5 bar assembly load, 19.7-mm-square chip, 200 W, temperature ~30°C.

Single-chip test vehicles were built to observe the effect of high-stress power cycles on the reliability of HNC and flat bondlines in a package configuration similar to those used in commercial products. The HNC modules were made by sawing the channels into the copper chip cap. Both HNC and flat modules were made with the same underfilled heater test chip with a ceramic chip carrier and the IBM ATC TIM. The test modules were stressed to 120°C junction temperature for 7 min and cooled to 20°C for 5 min with a 1-min gradual change for more than 2000 cycles, with the thermal resistance monitored between the chip center and cap. We observed a decrease in thermal resistance at the completion of 2000 cycles for HNC modules, compared with a more or less equal thermal resistance over time for flat modules.

IV. DISCUSSION AND OUTLOOK

Hierarchical channels were used to provide improved squeezing of paste and improved control of particle stacking during bondline formation. Particle stacking occurs in highly filled materials because of pressure gradients developing during squeeze flow over a rectangular surface, resulting in non-uniform interface properties and thick bondlines with a large thermal resistance. Nested surface channel designs are presented to create a uniform pressure drop as interface material flows across a rectangular surface. Compared with flat surfaces, reductions in thermal resistance by a factor of 2 to 3 times are demonstrated, with similar reductions in bondline thickness and assembly pressure. The results presented in this paper demonstrate a significant improvement in bondline properties with different interface materials and HNC: Tests with the IBM Advanced Thermal Compound resulted in 27-μm-thick bondlines with a thermal resistance of 8 $Kmm^2/W$, compared with 46 μm and 12 $Kmm^2/W$ for a flat surface with identical assembly conditions. Assembly pressures required for HNC were nearly two times lower than those for a flat surface. High-stress thermal cycling with hermetically sealed single-chip test modules with cap-integrated HNC demonstrated a decrease in thermal resistance during "burn-in" thermal cycling of over 2000 cycles compared with a slight increase in resistance for flat interfaces. To enable short-term application in many different industries, we are evaluating various processes such as metal molding, coining, and embossing to mass-fabricate HNC structures on chip caps. Many chip caps are already being created using such steps, allowing a potential low-cost HNC integration (< $1) when produced in high volumes.

Although the test results with existing pastes and greases already look promising, even larger benefits are expected from materials optimized for use with HNC. The simplest improvements would be to increase the matrix viscosity and TIM particle loading to enable improved conductivity rather than reduced bondline thickness. Because of the low implementation cost, an HNC chip cap could also be used for a purely mechanical benefit on lid-attach assembly lines, in which the ability to apply higher pressure loads is limited. Ongoing work suggests that the HNC cell size and hierarchy levels become the defining geometry of the TIM and may effectively de-couple the relationship between planar geometry and bondline thickness and could change the design points for future materials.

The ability to control particle stacking during bondline formation also opens up new possibilities for interface design. For several decades, particle-filled materials have been limited by percolation-based mixtures that deliver conductivities nearly an order of magnitude below the properties of the particles. With improved particle size and shape control during synthesis and further optimizations in the nested channel design, interfaces made from packed particle arrays with an effective conductivity closer to that of the particle material may be realized. Further research into the properties of the stacking line in comparison with surrounding bulk material is ongoing. The ability to control micro-particle position and orientation could also be extended to electrical contacts and nano-particle self-assembly so that dense 2D arrays of contacts and devices could be created.

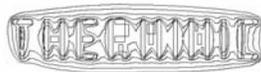

We have thoroughly demonstrated the benefit of HNC interfaces for small chip-to-cap thermal interfaces. Based on theoretical consideration and initial experiments we hypothesize that even larger interfaces such as cap-to-heatsink or interfaces for large power devices (IGBT, MOSFET, diodes) will benefit even more from HNC interfaces. This will allow improved thermal management in a wide range of applications and industries. With HNC technology, thermal interface thicknesses can be pushed to much lower values than flat interfaces so that filler-particle diameters can be reduced to the submicron scale as will be required for stacked or vertical IC integration. The thermal resistance in these applications will largely depend on the nanoscale structure of the interfaces and on the exact composition of chemicals at interfaces. Future work will thus need to focus on nanoscale thermal models such as molecular dynamics to understand these effects and to reduce these resistance components. Increased efforts in packaging are of great importance because a large thermal resistance between junction and ambient requires a large energy to cool the air and liquid used for convective cooling in a data center. New packages with reduced thermal resistance due to better interfaces and, possibly, due to direct liquid cooling can then reduce the required temperature difference needed for effective cooling and in turn the energy required for data center thermal management.

ACKNOWLEDGMENT

The authors thank Momentive Performance Materials and our colleagues, Hilton Toy, Rajneesh Kumar, Ijeoma Nnebe, and Steve Ostrander, for their efforts in providing sample materials. We also like to acknowledge Hugo Rothuizen, Urs Kloter, Mike Gaynes, John Magerlein, Shushumna Iruvanti, Kamal Sikka, and Paul Seidler for discussions and support.

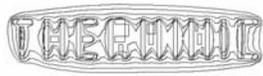